\documentclass[aps,prx,longbibliography,twocolumn,showpacs,preprintnumbers,floatfix,superscriptaddress]{revtex4-1}

\usepackage{graphicx}
\usepackage{dcolumn}
\usepackage{subfigure}
\usepackage{wrapfig}
\usepackage{cancel}
\usepackage{color}
\usepackage{textcomp}
\usepackage{amsmath}
\usepackage{amsfonts}
\usepackage{amssymb}
\usepackage{array}
\usepackage[colorlinks,urlcolor=blue,citecolor=blue,linkcolor=blue]{hyperref}

\newcommand{\cacuo}{Ca$_2$CuO$_3$}
\newcommand{\srcuo}{Sr$_2$CuO$_3$}

\newcommand{\ie}{\it{i.e.}}

\newcommand{\abinitio}{\it{ab initio}}

\begin{document}

\title{{\it Ab initio} quantum Monte Carlo calculations of spin superexchange
in cuprates:
the benchmarking case of {\cacuo}}
\author{Kateryna Foyevtsova}
\affiliation{Materials Science and Technology Division,
Oak Ridge National Laboratory, Oak Ridge, TN 37831, USA}
\author{Jaron T.~Krogel}
\affiliation{Materials Science and Technology Division,
Oak Ridge National Laboratory, Oak Ridge, TN 37831, USA}
\author{Jeongnim Kim}
\affiliation{Materials Science and Technology Division,
Oak Ridge National Laboratory, Oak Ridge, TN 37831, USA}
\author{P.~R.~C.~Kent}
\affiliation{Center for Nanophase Materials Sciences
and Computer Science and Mathematics Division,
Oak Ridge National Laboratory, Oak Ridge, TN 37831, USA}
\author{Elbio Dagotto}
\affiliation{Materials Science and Technology Division,
Oak Ridge National Laboratory, Oak Ridge, TN 37831, USA}
\affiliation{Department of Physics and Astronomy, The University of
Tennessee, Knoxville, TN 37996, USA}
\author{Fernando A.~Reboredo}
\affiliation{Materials Science and Technology Division,
Oak Ridge National Laboratory, Oak Ridge, TN 37831, USA}

\date{\today }

\begin{abstract}
In view of the continuous theoretical efforts
aimed at an accurate microscopic description of the strongly
correlated transition metal oxides and related materials,
we show that with continuum quantum Monte Carlo (QMC) calculations it is possible to
obtain the value of the spin superexchange coupling constant of a copper
oxide in a quantitatively
excellent agreement with experiment. The variational nature
of the QMC total energy allows us to identify the best
trial wave function out of the available pool of wave functions, which
makes the approach essentially free from adjustable parameters
and thus truly {\abinitio}. The present results on magnetic interactions
suggest that QMC is capable of accurately describing ground state
properties of strongly correlated materials.
\end{abstract}

\maketitle

For decades, transition metal oxides
have been amongst the most intriguing materials due to the complex correlated behavior
of the 3$d$ or 4$d$ electrons of a transition metal ion.
In particular, strong electronic correlations often give rise to non-trivial
magnetism, such as quantum spin-liquid states in low-dimensional Mott
insulating oxides. High-temperature superconductivity in copper oxides
(cuprates) is also believed to originate from magnetic spin excitations that bind
Cooper pairs \cite{Anderson87,Scalapino95,Dagotto94}. Electronic correlations, however, also make
this class of materials among the most difficult to describe theoretically, both from model as well as
{\abinitio} perspectives.

One of the practical challenges of
{\abinitio} electronic structure theory is to accurately
predict the strength of magnetic coupling between localized spins
of transition metal ions \cite{Janson12}. For solids, a natural method of choice is periodic density functional theory (DFT). DFT gives access to the system's ground
state energy corresponding to different configurations of localized spins,
which can be mapped onto the eigenstates of the a spin model to
extract the magnetic couplings $J$. This approach relies on the accuracy
of the description of the ground state. Unfortunately,
the presently available approximations to the exchange-correlation
functional in DFT either poorly account for the exchange and correlation effects
[local density approximation (LDA)] or depend on empirical input
parameters (LDA+U, hybrid functionals). Often, the only
way to find an appropriate approximation for the system of interest is
by comparing theoretical calculations with experiment,
compromising the predictive nature of such calculations.

\begin{table}[b]
\begin{tabular}{lr@{.}ll}
\hline\hline
\multicolumn{1}{>{\centering}p{0.45\columnwidth}<{\centering}}{Method} &
\multicolumn{3}{>{\centering}p{0.45\columnwidth}<{\centering}}{$J$ (eV)} \\
\hline
\hspace{0.8cm}Experiment [INS] 		& 0&241(11)&Ref.~\onlinecite{Walters09}\\  
\hspace{0.8cm}Experiment [$\chi(T)$]	& 0&146(13)&Refs.~\onlinecite{Ami95,Eggert96}\\
					& 0&189(17)&Ref.~\onlinecite{Motoyama96} \\
\hspace{0.8cm}FP-DMC 			& 0&159(14)&This work\\
                                        & 0&115(10)&This work \cite{BA}\\
\hspace{0.8cm}Cluster DDCI3 		& 0&231&Ref.~\onlinecite{Graaf00}\\
\hspace{0.8cm}UHF 			& 0&04 &Ref.~\onlinecite{Graaf00}\\
\hspace{0.8cm}LDA & \hspace{0.9cm}	  0&64 &This work\\
\hline\hline
\end{tabular}
\caption{
The nearest-neighbor spin superexchange coupling constant $J$
of {\cacuo} and {\srcuo} obtained with different
theoretical ({\cacuo}) and experimental ({\srcuo}) methods.
The abbreviations used stand for: 
INS=inelastic neutron scattering,
FP-DMC=fixed-phase diffusion Monte Carlo, 
DDCI3=difference dedicated configuration interaction with three degrees of freedom,UHF=unrestricted Hartree-Fock, 
LDA=local density approximation.
$\chi(T)$ denotes a temperature dependent magnetic susceptibility.
}
\label{T.Js}
\end{table}

\begin{figure}[tb]
\begin{center}
\includegraphics[trim = 0mm 0mm 0mm 0mm, clip,width=1.0\columnwidth]{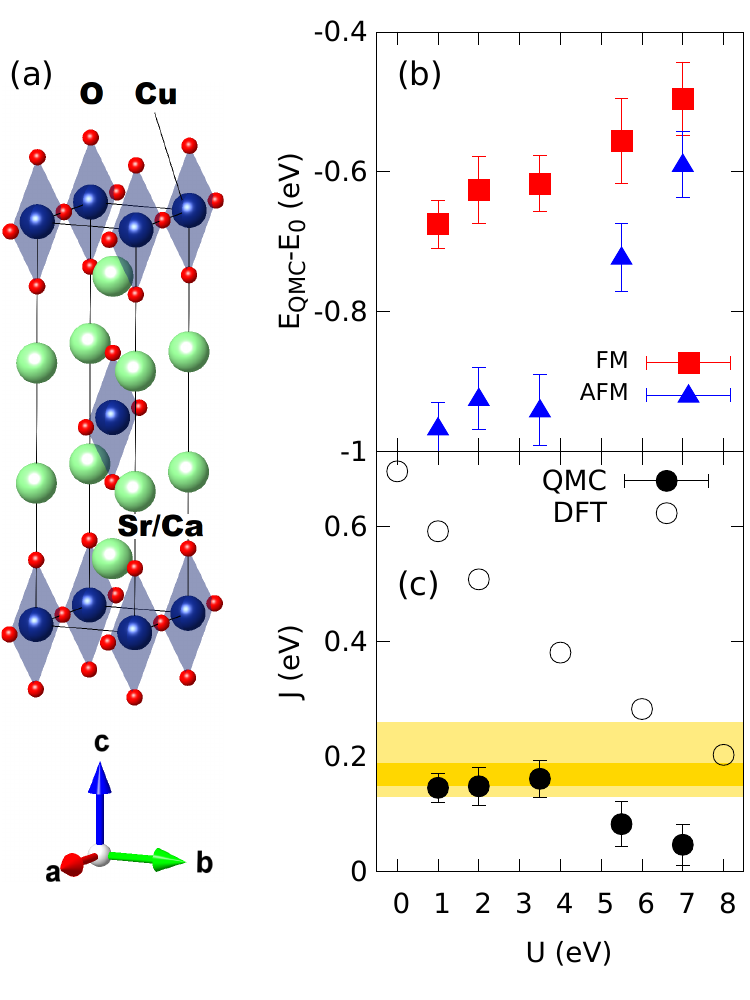}
\end{center}
\caption{(a) The conventional ``$1\times1\times1$'' unit cell of {\cacuo} and {\srcuo}.
For calculations, the unit cell crystallographic parameters
reported in Ref.~\onlinecite{Teske70} were used.
(b) The total energies of the antiferromagnetic
and ferromagnetic states of {\cacuo} calculated with FP-DMC
as a function of the trial wave function, characterized by
the LDA+U parameter $U$, used here as a variational parameter. For convenience of presentation the FP-DMC
energies are shifted by $E_0$=34705~eV.
(c) The nearest-neighbor Cu spin superexchange coupling constant $J$
of {\cacuo} calculated with WIEN2k (DFT) and FP-DMC as a function of $U$.
For comparison, the ranges of experimentally determined $J$
values of {\srcuo} are shown in light gold and dark gold bands.
The light gold band represents all reported experimental estimates \cite{expJ},
while the dark gold band represents the magnetic susceptibility measurements \cite{Eggert96,Motoyama96}.
}
\label{F.Js}
\end{figure}

This problem is generic to a broad class of transition metal oxides.
Let us exemplify the aforementioned limitations of DFT
by considering the case of the Mott insulators {\cacuo} and {\srcuo}.
These systems are one of the best realizations
of the one-dimensional (1D) spin-$\frac12$ antiferromagnetic Heisenberg chain model,
demonstrating spin-liquid behavior and separation of spin and orbital
degrees of freedom \cite{Schlappa12}.
The crystal structure of {\cacuo} and {\srcuo} is similar to that of
the superconducting two-dimensional cuprates, with the difference that
in the CuO$_2$ plane the oxygen atoms along the crystallographic $b$ direction are missing,
so that the Cu chains run along the $a$ direction [Fig.~\ref{F.Js}~(a)].
The Cu-O-Cu bridge provides a favorable path for superexchange coupling between Cu spins,
resulting in a particularly strong coupling constant $J$.
The experimental estimate of $J$ has been extracted from various probes,
performed mostly on {\srcuo}, and ranges between 0.13 and 0.26~eV (see
Table~\ref{T.Js} and Ref.~\onlinecite{expJ} for details).
Temperature dependent magnetic susceptibility measurements, supported by rigorous 
theoretical modeling  \cite{Eggert94,Eggert96,Motoyama96}, narrow this window to
0.15-0.19~eV.
Theoretical predictions of $J$, in turn, vary drastically by as much as an order of magnitude
depending on the method used. For example, periodic DFT with LDA gives 0.64~eV, whereas the periodic
unrestricted Hartree-Fock (UHF) method gives 0.04~eV
(Table~\ref{T.Js}). 
In addition, LDA+U calculations give results that depend strongly on $U$.
Cluster calculations with the configuration interaction method may give
a more reasonable result, as in this case (Table~\ref{T.Js},
Ref.~\onlinecite{Graaf00}), but,
generally speaking, their applicability to condensed phase systems is intrinsically limited.
In view of this, developing a new, more universal and accurate,
{\abinitio} approach to computing magnetic interactions
is critically important. Needless to say,
a method capable of accurately describing magnetic interactions
in transition metal oxides will also provide an improved {\abinitio} description of
many other ground state properties of these complex systems, and hence
may yield new physical insights.

Here, we apply
the diffusion Monte Carlo method \cite{foulkes01} within the fixed-phase approximation (FP-DMC)
to compute the value of the spin superexchange
interaction constant in a transition metal oxide.
To the best of our knowledge, this is among the first calculations
of magnetic couplings in complex oxides that has been attempted with 
a method capable of chemical accuracy in full periodic boundary conditions.
The fixed-phase error is controlled by scanning over a set of trial wave functions
and using the variational nature of DMC energy, as explained below.
This way, the variational principle determines the choice of the initial
DFT functional to generate a trial wave function and thus
eliminates empiricism from the calculations.
We choose the 1D cuprate antiferromagnet {\cacuo}
as our test system because of the simplicity of its
underlying spin model and relatively light constituent atoms,
as compared to other cuprates, including {\srcuo}, to minimize the
potential role of
relativistic effects.
Our result for the nearest-neighbor Cu spin coupling, $J=0.159\pm0.014$~eV \cite{BA},
is in excellent agreement
with the value extracted from the temperature dependence of
magnetic susceptibility (Refs.~\onlinecite{Ami95,Eggert96,Motoyama96}, Table~\ref{T.Js}).

Historically, quantum Monte Carlo has played a fundamental role
in advancing electronic structure theory.
Released-node DMC calculations on the homogeneous electron gas
by Ceperley and Alder \cite{Ceperley80} were used to construct the
local density approximation to the exchange-correlation functional
which is at the core of modern DFT methods. 
In FP-DMC, the Schr\"{o}dinger equation is rewritten in a form
of an integral diffusion equation, which is stochastically
solved via quantum Monte Carlo sampling by iteratively propagating the
wave function
in imaginary time. As a result, the ground state (GS) wave function is projected out.
In order to handle the fermion sign problem, the complex phase
of the target GS wave function is fixed to those of an input trial wave function.
The fixed-phase approximation introduces a variational systematic
error in the DMC energy which can
be assessed and controlled by comparing results obtained with
different trial wave functions.
Exchange and correlation effects are fully accounted for within the fixed phase approximation.
Even though the computational costs of FP-DMC grow as $N^3$, where $N$ is
the number of particles, the costs of calculating quantities {\it per particle}
grow only as $N^2$, which is a great advantage in the present case since
$J$ is a characteristic of a {\it Cu-Cu bond}. Due to improvements in
algorithms and available computer power, QMC has achieved a growing
success in accurately predicting the properties of complex
materials \cite{shulenb13,booth13,hood12,esler10,spanu09,kolorenc08}.

We used the DFT plane-wave code Quantum Espresso  \cite{qe}
in order to self-consistently
generate trial wave functions in a single Slater determinant form.
The LDA+U method was applied, with the values of the on-site
Coulomb repulsion between Cu $3d$ electrons, $U$, varying as 1, 2,
3.5, 5.5, and 7~eV. The energy cut-off was set to 500~Ry due to the
use of very hard pseudopotentials and inclusion of semi-core electrons
in the valence (described below). The corresponding DMC energies were subsequently
compared to determine the best trial wave function.
We present results that follow procedures in the literature \cite{mazin08,ma08,akamatsu11,seo12,duan06,reinhardt99}
to obtain the spin superexchange coupling constant $J$.
The coupling constant can be computed from a single total energy difference: 
$J = (E_{FM}-E_{AFM})/(2N_{Cu-Cu}s_z^2)$.
Here $s_z=1/2$ is the z-component of an electron's spin and $N_{Cu-Cu}$ 
is the number of nearest neighbor Cu-Cu bonds in a given supercell. 
This results from a mapping to the total energy differences of an 
Ising Hamiltonian.  The estimate for $J$ changes somewhat using a different 
approach \cite{BA}, though the result remains close to the experimental range.
Although the total energies are variational, $J$ is not and care must be 
taken to fully optimize the trial wavefunctions.
Wavefunctions of the standard spin-assigned Slater-Jastrow type \cite{foulkes01} 
used here are eigenstates of total $\hat{S}_z$, but not $\hat{S}^2$. 
Although it is possible to construct eigenstates of $\hat{S}^2$, this is seldom 
done in practice with DMC because the Slater-Jastrow form gives accurate total 
energies within the fixed-node/phase approximation \cite{ceperley91} 
for spin-independent Hamiltonians.
For all considered $U$ values, both FM and AFM solutions are insulating in DFT.

Fixed phase DMC calculations were performed with QMCPACK  \cite{qmcpack}.
The DMC imaginary time step and the number of walkers have been converged to,
respectively, 0.005~Ha$^{-1}$ and 2000 per boundary twist (in the $2\times1\times1$ supercell).
To assess finite size errors we considered $2\times1\times1$ ($N_{Cu-Cu}=4$) and
$2\times2\times1$ ($N_{Cu-Cu}=8$) supercells, defined with respect to the conventional $Immm$ crystallographic
unit cell of {\cacuo} [Fig.~\ref{F.Js}~(a)].
The $2\times1\times1$ supercell contains four formula units, {\ie}, 28 atoms with 228 electrons.
We also performed averaging over twisted boundary conditions on a $2\times4\times1$ $k$-point grid
for the $2\times1\times1$ supercell ($2\times2\times1$ for the $2\times2\times1$ supercell).
The necessity of twist averaging indicates that cluster model
calculations, such as
those in Ref. \onlinecite{Graaf00}, are under converged with respect to finite size effects.  
The ionic potentials were approximated
by employing pseudopotentials (PPs).  
Through an extensive investigation, we have found the inclusion of semi-core electrons
to be essential for high quality results.  Core sizes used for the pseudopotentials
are as follows:
 He-core for oxygen atoms (6 electrons in valence),
Ne-core for calcium atoms (10 electrons in valence), and Ne-core
for copper atoms (19 electrons in valence).
The quality of the PPs has been carefully tested within both DFT and DMC,
as reported in detail in Appendix \ref{app}.

We first present the FP-DMC results obtained for the $2\times1\times1$ supercell,
which contains two Cu atoms along the chain direction $a$.
Figure~\ref{F.Js}~(b) displays the FP-DMC total energies
of the FM and AFM states as a function of the trial wave function,
characterized by the LDA+U parameter $U$. In these calculations we stress that the U is
simply a convenient optimization parameter for generating FP-DMC wave functions.
Both the FM and AFM curves follow a non-linear $U$ dependence,
reaching minima and leveling off in the region between $U$=1 and 3.5~eV, within
the available statistical resolution.

From the difference between  the AFM and FM FP-DMC total energies
we compute the spin superexchange constant $J$, shown in Fig.~\ref{F.Js}~(c)
together with the respective LDA+U $J$ values for comparison.
Also indicated are the ranges of experimentally determined $J$ values of {\srcuo}:
the light gold band represents all reported experimental estimates \cite{expJ}, while
the dark gold band represents susceptibility measurements \cite{Ami95,Eggert96,Motoyama96},
which is one of the most reliable probes.
Since, unfortunately, no equivalent experimental data on {\cacuo}
are available, we can only compare our theoretical calculations
with the experiments performed on {\srcuo}.
This is a valid approach as the spin exchange couplings
of the two cuprates should differ by no more than a few percent \cite{Rosner99}.
From Fig.~\ref{F.Js}~(c), one readily sees that
in the $U$ region between 1 and 3.5~eV,
corresponding to the minimal FP-DMC energies
in Fig.~\ref{F.Js}~(b),
the FP-DMC results for $J$
are in good agreement with the susceptibility data,
within statistical resolution.
In contrast, all electron LDA+U (LAPW) results strongly depend on $U$, 
requiring a large value of $U\approx 8$ to obtain reasonable $J$-values. 

We now assess the finite size error associated with these results
by performing FP-DMC calculations on a $2\times2\times1$ supercell,
obtained by doubling the original $2\times1\times1$ supercell
in the direction perpendicular to the Cu chains.
The $U$=3.5~eV LDA+U trial wave function is used here
as the one to provide a good
complex phase for FP-DMC,
as has been established above. The resulting $J=0.159(14)$~eV
is to be compared with the $2\times1\times1$ result of $0.16(3)$~eV.
From this, we conclude that the finite size error must be within $0.03$~eV,
which is the statistical accuracy of the $2\times1\times1$ calculations
of Fig.~\ref{F.Js}~(c).

We would also like to give a brief comment here on the computational
costs involved. The $J$ values for 2~eV $<U<$ 7~eV in Fig.~3~(c) cost
$\sim$100-130K cpu hours each (error bar $\sim$0.045 eV).
Calculating $J$ for the $2\times2\times1$
supercell took 1.8M cpu hours. We note that such costs are not 
insignificant, but are affordable 
on modern supercomputers such as Titan at ORNL.

In conclusion, we have presented a
theoretical determination of the value
of the spin superexchange constant in a transition
metal oxide with the FP-DMC method.
Our results for the 1D antiferromagnetic
cuprate {\cacuo} are in excellent agreement with experiment.
Moreover, this is a purely {\abinitio} approach, where
the fixed phase error is controlled via the variational principle,
with no empirical adjustable parameters. In this sense,
FP-DMC is superior to DFT where in order to
improve the description of exchange and correlations one
often resorts to LDA+U or hybrid functionals and chooses the ``best''
functional empirically.
The success of FP-DMC in the present case
implies that this method is capable of accurately describing
the complicated spin superexchange processes between
the correlated Cu $3d$ orbitals and oxygen $2p$ orbitals,
involving on-site Coulomb correlations and $p-d$ orbital
hybridization.
We hope that our present successful application of FP-DMC
will stimulate future studies of
magnetic and other properties of strongly correlated transition metal oxides
with this highly competitive {\abinitio} method.

Note: At the time of submitting this manuscript, we learned of similar DMC calculations, 
performed independently, published on \texttt{arXiv.org}  \cite{Wagner13}. 

The authors would like to thank L. Shulenburger for sharing expertise in 
pseudopotential construction and for providing access to pseudopotential 
datasets prior to publication.
The work was supported by the Materials Sciences \& Engineering Division of the
Office of Basic Energy Sciences, U.S. Department of Energy. PRCK
was supported by the Scientific User Facilities Division, Office of
Basic Energy Sciences, U.S. Department of Energy. Computational time
used resources of the Oak Ridge Leadership Computing Facility at the
Oak Ridge National Laboratory, which is supported by the Office of
Science of the U.S. Department of Energy under Contract
No. DE-AC05-00OR22725.

\appendix
\section{Pseudopotential tests} \label{app}

For oxygen and calcium we used the pseudopotentials (PPs) optimized
by Shulenburger and Mattsson \cite{shulenb13}, who
also demonstrated their good quality by performing numerous tests.
Of a much greater concern in the present study was the proper performance of the Cu PP
since the magnetic properties of {\cacuo} are largely determined by
the behavior of the Cu $3d$ electrons.
Therefore, we subjected our candidate Cu PPs to a comprehensive selection
process, as presented below. The candidate Cu PPs were
generated using the Opium code \cite{opium}. Our final choice is the hard Ne-core
Cu PP [Fig.~\ref{F.Ne_HardSoft}~(a), (c)] that satisfies the most stringent accuracy criteria and thus
ensures the validity of the bulk calculations.

\subsection{Bulk DFT calculations and rejection of Ar-core Cu pseudopotentials}
\begin{figure}[tb]
\begin{center}
\includegraphics[trim = 0mm 0mm 0mm 0mm, clip,width=1.0\columnwidth]{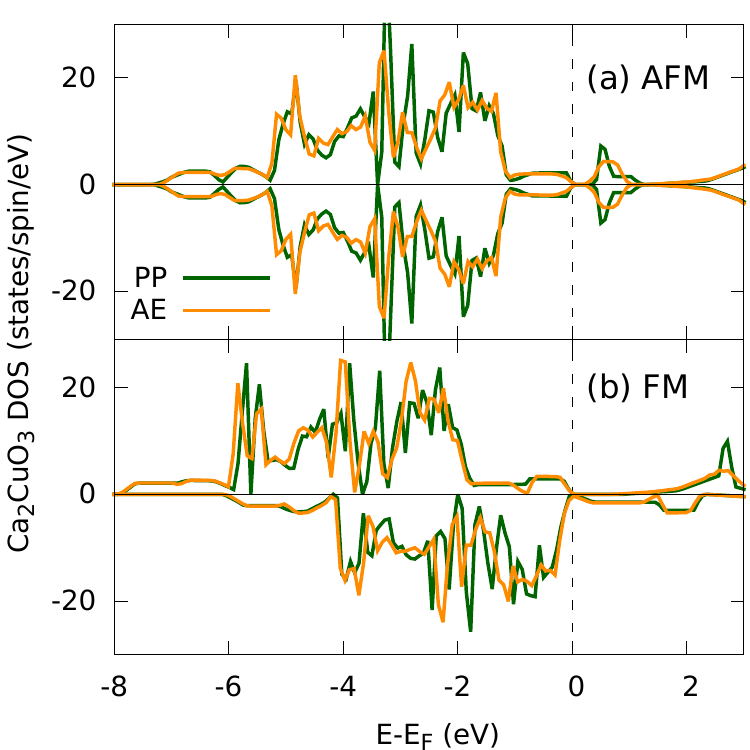}
\end{center}
\caption{The comparison of the {\cacuo}
densities of states (DOS) calculated using PP and AE codes:
(a) antiferromagnetic and (b) ferromagnetic states.
Energies are measured relative to the Fermi level $E_{\rm F}$}
\label{F.dos}
\end{figure}

\begin{figure*}[tb]
\begin{center}
\includegraphics[trim = 0mm 0mm 0mm 0mm, clip,width=2.0\columnwidth]{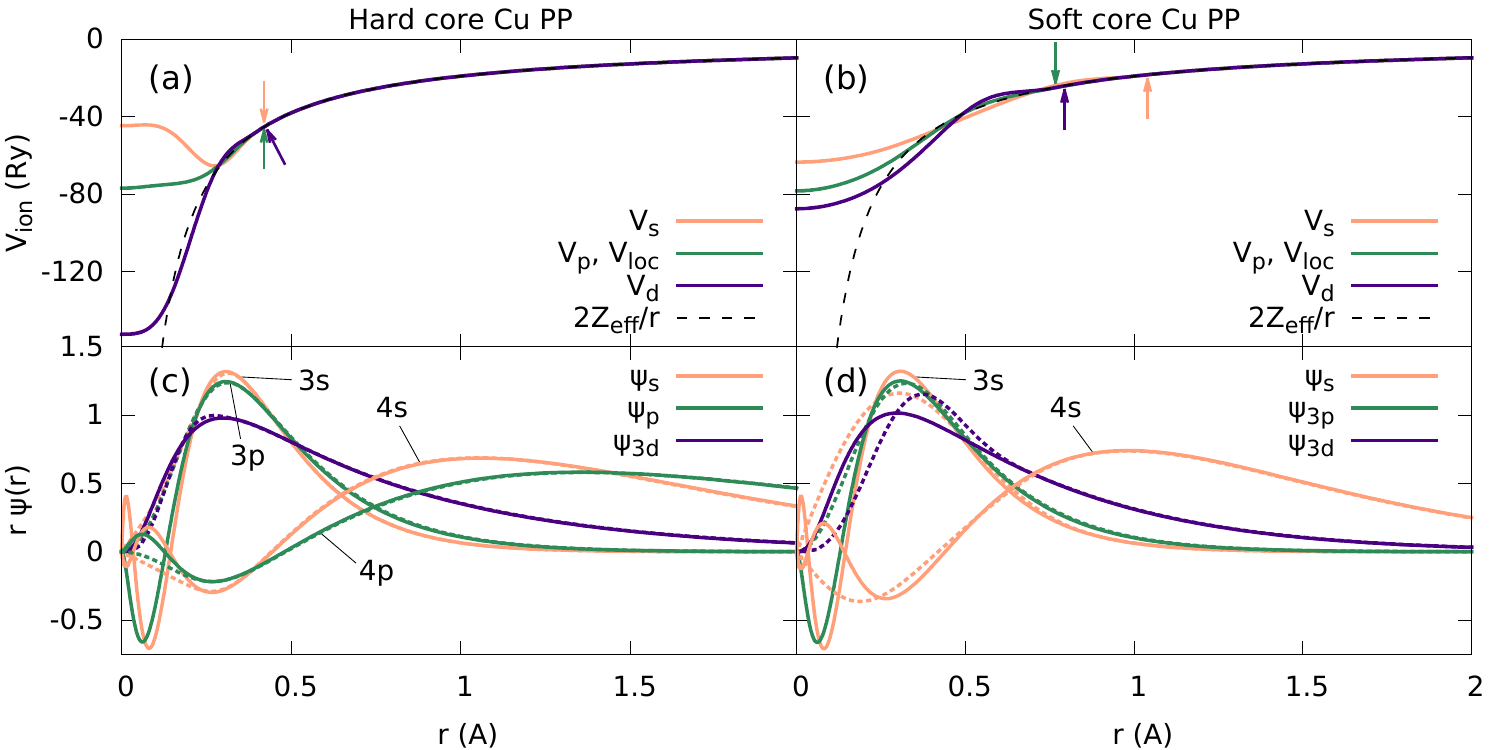}
\end{center}
\caption{(a) Hard and (b) soft Ne-core Cu pseudopotentials (PPs).
$V_{\text{loc}}$, $V_s$, $V_p$, and $V_d$ represent
the local, $s$-, $p$-, and $d$-channels, respectively. $2Z_{\text{eff}}/r$
represents the Coulomb potential due to the effective charge $Z_{\text{eff}}$.
The arrowed lines of respective colors indicate
cut-off radii for the three pseudopotential channels.
(c), (d): The Cu atom eigenfunctions obtained in LDA with the hard (c)
and soft (d) Ne-core Cu PPs. The solid (dashed) lines represent wave functions
obtained from all-electron (PP) calculations.
}
\label{F.Ne_HardSoft}
\end{figure*}

Using our hard Ne-core Cu PP, we are able to accurately reproduce
with Quantum Espresso the LDA results of the all-electron (AE)
code WIEN2k \cite{wien2k} for bulk {\cacuo} in 
an antiferromagnetic (AFM) and a ferromagnetic (FM) states.
Thus, in Fig.~\ref{F.dos} the {\cacuo}
densities of states (DOS) obtained from PP and AE calculations are compared,
for an AFM [Fig.~\ref{F.dos}~(a)] and an FM [Fig.~\ref{F.dos}~(b)] Cu spin configurations.
For both configurations, the agreement with the AE code is very good.
Interestingly, we were able to equally well reproduce the AE DOS
when also using properly optimized Mg-core Cu PPs. This, however,
did not hold for any of the Ar-core Cu PPs we tried: in this case,
the bandwidth of the PP states as well as the conduction
gap (AFM configuration) are systematically larger than in the AE calculations.
This allowed us to discard Ar-core Cu PPs already at this stage of
PP validation.

As for the spin superexchange coupling constant $J$,
Quantum Espresso with the hard Ne-core Cu PP gives 0.64~eV in LDA,
while WIEN2k gives 0.72~eV. 
In LDA+U, the WIEN2k $J$ is rapidly decreasing
as $1/U$.  In PP DFT, we find only a weak dependence on $U$.
This may be a peculiarity of the implementation of the LDA+U scheme within the
plane-wave basis method of Quantum Espresso.

\subsection{FP-DMC atom ionization energies and rejection of Mg-core Cu pseudopotentials}
As we have pointed out, in DFT calculations for bulk {\cacuo},
the Ne-core and the Mg-core Cu PPs appear to be equally good,
provided proper optimization has been carried out.
However, this does not necessarily mean that they
will perform well in diffusion Monte Carlo calculations.
In order to test the latter, we
calculated the Cu atom ionization energy with FP-DMC,
using LDA for generating trial wave functions.
The FP-DMC computational parameters are the following:
$3.125\cdot10^{-4}$~Ha$^{-1}$ for the imaginary time step and
4000 for the number of DMC walkers.
With the hard Ne-core and with the Mg-core Cu PPs,
we obtain, respectively, 7.724(37)~eV and 8.302(36)~eV
for the Cu atom ionization energy. The former
number is in a much better agreement with the
experimental result of 7.72638(1)~eV.
One of the reasons of the poor performance of the Mg-core
PP in this test, is that the $3s$ orbital, which is
treated as core here, has a significant overlap in space
with the $3p$ orbital, treated as valence [see Fig.~\ref{F.Ne_HardSoft}~(c)].
This causes less trouble in DFT which is formulated in terms
of Kohn-Sham orbitals so that such a division based on orbital character is natural.
DMC, on the other hand, operates with a full many-electron wave function
where a removal of the $3s$ electrons negatively affects
the representation of the motion of the nearby $3p$ electrons.
This issue has also been discussed in the context of GW  \cite{Dixit10,Dixit11}.

We would like to point out that
using a Mg-core Cu PP instead of a Ne-core one
in
bulk {\cacuo} FP-DMC calculations
could provide a speedup of more than 30\%,
owing to the fact that the deeply lying $3s$ electrons
are a significant source of the energy variance.

\subsection{Equation of state of CuO dimer and hard vs. soft Ne-core
Cu pseudopotentials}
\begin{figure}[tb]
\begin{center}
\includegraphics[trim = 0mm 0mm 0mm 0mm, clip,width=1.0\columnwidth]{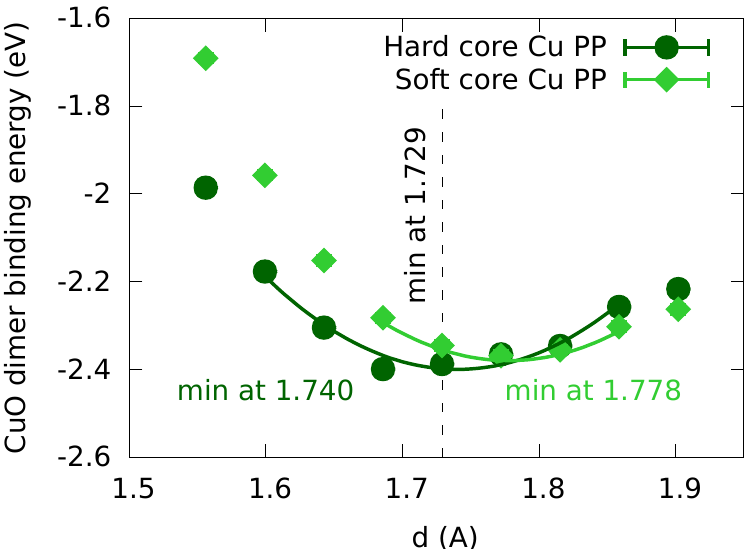}
\end{center}
\caption{Binding energy as a function of the
interatomic separation $d$ in a CuO dimer molecule.
The vertical dashed line indicates the experimental equilibrium
separation. The solid lines are a quadratic fit to the data. }
\label{F.cuo_eos}
\end{figure}

Although the hard Ne-core PP has been proven to be of a good quality
for both DFT and DMC, it has a disadvantage in terms of computational
load and memory demands as it requires a minimum of 500~Ry energy cut-off for the Quantum
Espresso plane-wave basis. This results from the quite small cut-off radii
of the $s$-, $p$-, and $d$-channels, as shown in Fig.~\ref{F.Ne_HardSoft}~(a).
In view of this, we constructed and tested an alternative
Ne-core Cu PP with a soft core and low energy cut-off requirements of less that 200~Ry
[Fig.~\ref{F.Ne_HardSoft}~(b)]. It demonstrated excellent
characteristics in DFT tests but, unfortunately, gives
worse results in DMC than the hard-core Cu PP.
In particular, with the soft-core PP the equilibrium
interatomic separation distance in a CuO molecule
is overestimated by more than 3\% in FP-DMC
[holds for LDA, LDA+U ($U$=3.5, 6~eV), and B3LYP wave functions],
whereas with the hard-core PP
it is overestimated by only 0.6\% (Fig.~\ref{F.cuo_eos}).
Also the Cu atom ionization energy is slightly underestimated:
7.548(42)~eV. Thus all calculations in DMC reported in the paper were
obtained with the hard Ne-core pseudopotential. The surprising
sensitivities to pseudopotential formulation that we find even for
small core pseudopotentials indicates that careful testing is
essential and results from large core pseudopotentials must be treated
with caution.

\newpage

%\bibliography{ref}

\begin{thebibliography}{}

\bibitem{Anderson87}
P.~W.~Anderson,
\emph{The Resonating Valence Bond State in La2CuO4 and Superconductivity},
\href{http://dx.doi.org/10.1126/science.235.4793.1196}{Science {\bf 235}, 1196 (1987)}.

\bibitem{Scalapino95}
D.~Scalapino,
\emph{The case for $d_{x^2 - y^2}$ pairing in the cuprate superconductors},
\href{http://dx.doi.org/10.1016/0370-1573(94)00086-I}{Phys. Rep. {\bf 250}, 329 (1995)}.

\bibitem{Dagotto94}
E.~Dagotto,
\emph{Correlated electrons in high-temperature superconductors},
\href{http://dx.doi.org/10.1103/RevModPhys.66.763}{Rev. Mod. Phys. {\bf 66}, 763 (1994)}.

\bibitem{Janson12}
O.~Janson,
\emph{D cuprate microscopic magnetic modeling for low-dimensional spin systems},
Ph.D. thesis, Technische Universit\"{a}t Dresden, (2012)

\bibitem{Walters09}
A.~C.~Walters, T.~G.~Perring, J.~Caux, A.~T.~Savici, G.~D.~Gu, C.~Lee, W.~Ku, and I.~A.~Zaliznyak,
\emph{Effect of covalent bonding on magnetism and the missing neutron intensity	in copper oxide compounds},
\href{http://dx.doi.org/10.1038/nphys1405}{Nat. Phys. {\bf 5}, 867 (2009)}.

\bibitem{Ami95}
T.~Ami, M.~K.~Crawford, R.~L.~Harlow, Z.~R.~Wang, D.~C.~Johnston, Q.~Huang, and R.~W.~Erwin,
\emph{Magnetic susceptibility and low-temperature structure of the linear	chain cuprate ${\mathrm{Sr}}_{2}$${\mathrm{CuO}}_{3}$},
\href{http://dx.doi.org/10.1103/PhysRevB.51.5994}{Phys. Rev. B {\bf 51}, 5994 (1995)}.

\bibitem{Eggert96}
S.~Eggert,
\emph{Accurate determination of the exchange constant in ${\mathrm{Sr}}_{2}$${\mathrm{CuO}}_{3}$	from recent theoretical results},
\href{http://dx.doi.org/10.1103/PhysRevB.53.5116}{Phys. Rev. B {\bf 53}, 5116 (1996)}.

\bibitem{Motoyama96}
N.~Motoyama, H.~Eisaki, and S.~Uchida,
\emph{Magnetic Susceptibility of Ideal Spin 1$/$2 Heisenberg Antiferromagnetic	Chain Systems, ${\mathrm{Sr}}_{2}{\mathrm{CuO}}_{3}$ and ${\mathrm{SrCuO}}_{2}$},
\href{http://dx.doi.org/10.1103/PhysRevLett.76.3212}{Phys. Rev. Lett. {\bf 76}, 3212 (1996)}.

\bibitem{BA}
Associating the QMC AFM state with the Bethe anzatz ground-state 
of the 1D Heisenberg model (instead of the N'{e}el/Ising state as is usually done \cite{mazin08,ma08,akamatsu11,seo12,duan06,reinhardt99})  
gives a different representation of $J$: $J \approx 0.72 J^{Neel} = 0.115(10)$~eV.

\bibitem{Graaf00}
C.~de~Graaf and F.~Illas,
\emph{Electronic structure and magnetic interactions of the spin-chain	compounds ${\mathrm{Ca}}_{2}{\mathrm{CuO}}_{3}$ and ${\mathrm{Sr}}_{2}{\mathrm{CuO}}_{3}$},
\href{http://dx.doi.org/10.1103/PhysRevB.63.014404}{Phys. Rev. B {\bf 63}, 014404 (2000)}.

\bibitem{Schlappa12}
J.~Schlappa, K.~Wohlfeld, K.~J.~Zhou, M.~Mourigal, M.~W.~Haverkort, V.~N.~Strocov, L.~Hozoi, C.~Monney, S.~Nishimoto, S.~Singh, A.~Revcolevschi, J.~Caux, L.~Patthey, H.~M.~Rønnow, J.~van~den~Brink, and T.~Schmitt,
\emph{Spin–orbital separation in the quasi-one-dimensional Mott insulator	Sr$_2$CuO$_3$},
\href{http://dx.doi.org/10.1038/nature10974}{Nature {\bf 485}, 82 (2012)}.


\bibitem{expJ}
From the fit to the temperature dependence of magnetic susceptibility, 
the superexchange constant $J$ between the nearest-neighbor Cu spins 
in Sr$_2$CuO$_3$ was estimated as 0.146(13) \cite{Eggert96} and 0.189(17) \cite{Motoyama96}~eV. 
For the same compound, analysis of the angle-resolved photoemission 
spectroscopy data 
leads to a $J$ value between 0.13 and 0.16~eV \cite{Fujisawa98} while the 
midinfrared spectrum interpretation in terms of phonon-assisted magnetic 
excitations 
to a value between 0.246 and 0.26~eV \cite{Lorenzana97,Suzuura96}. 
Resonant inelastic 
x-ray scattering \cite{Schlappa12} and inelastic neutron scattering \cite{Walters09} 
data point to 0.249 and 0.241(11)~eV, respectively. Finally, from
the $^{63}$Cu neutron 
magnetic resonance data one infers the $J$ value of 0.24~eV \cite{Takigawa97,Takigawa96}.

\bibitem{Eggert94}
S.~Eggert, I.~Affleck, and M.~Takahashi,
\emph{Susceptibility of the spin 1/2 Heisenberg antiferromagnetic chain},
\href{http://dx.doi.org/10.1103/PhysRevLett.73.332}{Phys. Rev. Lett. {\bf 73}, 332 (1994)}.

\bibitem{foulkes01}
W.~M.~C.~Foulkes, L.~Mitas, R.~J.~Needs, and G.~Rajagopal,
\emph{Quantum Monte Carlo simulations of solids},
\href{http://dx.doi.org/10.1103/RevModPhys.73.33}{Rev. of Mod. Phys. {\bf 73}, 33 (2001)}.

\bibitem{Teske70}
C.~L.~Teske and H.~Müller-Buschbaum,
\emph{\"{U}ber Erdalkalimetall-Oxocuprate. IV. Zur Kenntnis von SrCu$_2$O$_2$},
\href{http://dx.doi.org/10.1002/zaac.19703790202}{Z. Anorg. Allg. Chem. {\bf 379}, 113 (1970)}.

\bibitem{Ceperley80}
D.~M.~Ceperley and B.~J.~Alder,
\emph{Ground State of the Electron Gas by a Stochastic Method},
\href{http://dx.doi.org/10.1103/PhysRevLett.45.566}{Phys. Rev. Lett. {\bf 45}, 566 (1980)}.

\bibitem{shulenb13}
L.~Shulenburger and T.~R.~Mattsson,
\emph{Quantum Monte Carlo applied to solids},
\href{http://dx.doi.org/10.1103/PhysRevB.88.245117}{Phys. Rev. B {\bf 88}, 245117 (2013)}.

\bibitem{booth13}
G.~H.~Booth, A.~Gr{\"{u}}neis, G.~Kresse, and A.~Alavi,
\emph{Towards an exact description of electronic wavefunctions in real solids},
\href{http://dx.doi.org/10.1038/nature11770}{Nature {\bf 493}, 365 (2013)}.

\bibitem{hood12}
R.~Q.~Hood, P.~R.~C.~Kent, and F.~A.~Reboredo,
\emph{Diffusion quantum Monte Carlo study of the equation of state and point defects in aluminum},
\href{http://dx.doi.org/10.1103/PhysRevB.85.134109}{Phys. Rev. B {\bf 85}, 134109 (2012)}.

\bibitem{esler10}
K.~P.~Esler, R.~E.~Cohen, B.~Militzer, J.~Kim, R.~J.~Needs, and M.~D.~Towler,
\emph{Fundamental High-Pressure Calibration from All-Electron Quantum Monte Carlo Calculations},
\href{http://dx.doi.org/10.1103/PhysRevLett.104.185702}{Phys. Rev. Lett. {\bf 104}, 185702 (2010)}.

\bibitem{spanu09}
L.~Spanu, S.~Sorella, and G.~Galli,
\emph{Nature and Strength of Interlayer Binding in Graphite},
\href{http://dx.doi.org/10.1103/PhysRevLett.103.196401}{Phys. Rev. Lett. {\bf 103}, 196401 (2009)}.

\bibitem{kolorenc08}
J.~Koloren\v{c} and L.~Mitas,
\emph{Quantum Monte Carlo Calculations of Structural Properties of FeO Under Pressure},
\href{http://dx.doi.org/10.1103/PhysRevLett.101.185502}{Phys. Rev. Lett. {\bf 101}, 185502 (2008)}.

\bibitem{qe}
P.~Giannozzi, S.~Baroni, N.~Bonini, M.~Calandra, R.~Car, C.~Cavazzoni, D.~Ceresoli, G.~L.~Chiarotti, M.~Cococcioni, I.~Dabo, A.~Dal~Corso, S.~de~Gironcoli, S.~Fabris, G.~Fratesi, R.~Gebauer, U.~Gerstmann, C.~Gougoussis, A.~Kokalj, M.~Lazzeri, L.~Martin-Samos, N.~Marzari, F.~Mauri, R.~Mazzarello, S.~Paolini, A.~Pasquarello, L.~Paulatto, C.~Sbraccia, S.~Scandolo, G.~Sclauzero, A.~P.~Seitsonen, A.~Smogunov, P.~Umari, and R.~M.~Wentzcovitch,
\emph{QUANTUM ESPRESSO: a modular and open-source software project for quantum simulations of materials},
\href{http://dx.doi.org/10.1088/0953-8984/21/39/395502}{J. Phys.: Condens. Matter {\bf 21}, 395502 (2009)}.

\bibitem{mazin08}
I.~I.~Mazin, M.~D.~Johannes, L.~Boeri, K.~Koepernik, and D.~J.~Singh,
\emph{Problems with reconciling density functional theory calculations with experiment in ferropnictides},
\href{http://dx.doi.org/10.1103/PhysRevB.78.085104}{Phys. Rev. B {\bf 78}, 085104 (2008)}.

\bibitem{ma08}
F.~Ma, Z.~Lu, and T.~Xiang,
\emph{Arsenic-bridged antiferromagnetic superexchange interactions in LaFeAsO},
\href{http://dx.doi.org/10.1103/PhysRevB.78.224517}{Phys. Rev. B {\bf 78}, 224517 (2008)}.

\bibitem{akamatsu11}
H.~Akamatsu, Y.~Kumagai, F.~Oba, K.~Fujita, H.~Murakami, K.~Tanaka, and I.~Tanaka,
\emph{Antiferromagnetic superexchange via 3$d$ states of titanium in EuTiO$_3$ as seen from hybrid Hartree-Fock density functional calculations},
\href{http://dx.doi.org/10.1103/PhysRevB.83.214421}{Phys. Rev. B {\bf 83}, 214421 (2011)}.

\bibitem{seo12}
H.~Seo, A.~Posadas, and A.~A.~Demkov,
\emph{Strain-driven spin-state transition and superexchange interaction in LaCoO$_3$: {\it Ab initio} study},
\href{http://dx.doi.org/10.1103/PhysRevB.86.014430}{Phys. Rev. B {\bf 86}, 014430 (2012)}.

\bibitem{duan06}
C.~Duan, R.~F.~Sabiryanov, W.~N.~Mei, P.~A.~Dowben, S.~S.~Jaswal, and E.~Y.~Tsymbal,
\emph{Magnetic ordering in Gd monopnictides: Indirect exchange versus superexchange interaction},
\href{http://dx.doi.org/10.1063/1.2200767}{Appl. Phys. Lett. {\bf 88},  (2006)}.

\bibitem{reinhardt99}
P.~Reinhardt, M.~P.~Habas, R.~Dovesi, I.~de~P.~R.~Moreira, and F.~Illas,
\emph{Magnetic coupling in the weak ferromagnet CuF$_2$},
\href{http://dx.doi.org/10.1103/PhysRevB.59.1016}{Phys. Rev. B {\bf 59}, 1016 (1999)}.

\bibitem{ceperley91}
D.~Ceperley,
\emph{Fermion nodes},
\href{http://dx.doi.org/10.1007/BF01030009}{J. Stat. Phys. {\bf 63}, 1237 (1991)}.

\bibitem{qmcpack}
J.~Kim, K.~P.~Esler, J.~McMinis, M.~A.~Morales, B.~K.~Clark, L.~Shulenburger, and D.~M.~Ceperley,
\emph{Hybrid algorithms in quantum Monte Carlo},
\href{http://dx.doi.org/10.1088/1742-6596/402/1/012008}{J. Phys. Conf. Ser. {\bf 402}, 012008 (2012)}.

\bibitem{Rosner99}
H.~Rosner,
\emph{Electronic structure and exchange integrals of low-dimensional cuprates},
Ph.D. thesis, Technische Universit\"{a}t Dresden, (1999)

\bibitem{Wagner13}
L.~K.~Wagner and P.~Abbamonte,
\emph{The effect of electron correlation on the electronic structure and spin-lattice coupling of the high-T$_c$ cuprates: quantum Monte Carlo calculations},
\href{http://www.arxiv.org/abs/1402.4680}{arXiv:1402.4680 (2014)}.

\bibitem{opium}
E.~Walter,
\emph{OPIUM},
\href{http://opium.sourceforge.net}{http://opium.sourceforge.net}

\bibitem{wien2k}
P.~ Blaha, K.~Schwarz, G.~K.~H.~Madsen, D.~ Kvasnicka, and J.~Luitz,
\emph{{WIEN2K}, {A}n {A}ugmented {P}lane {W}ave + {L}ocal {O}rbitals {P}rogram for {C}alculating {C}rystal {P}roperties},
{K}arlheinz Schwarz, Techn. Universit\"{a}t Wien, Austria, (2001)

\bibitem{Dixit10}
H.~Dixit, R.~Saniz, D.~Lamoen, and B.~Partoens,
\emph{The quasiparticle band structure of zincblende and rocksalt ZnO},
\href{http://dx.doi.org/10.1088/0953-8984/22/12/125505}{J. Phys.: Condens. Matter {\bf 22}, 125505 (2010)}.

\bibitem{Dixit11}
H.~Dixit, R.~Saniz, D.~Lamoen, and B.~Partoens,
\emph{Accurate pseudopotential description of the GW bandstructure of ZnO},
\href{http://dx.doi.org/10.1016/j.cpc.2011.02.001}{Comput. Phys. Commun. {\bf 182}, 2029 (2011)}.

\bibitem{Fujisawa98}
H.~Fujisawa, T.~Yokoya, T.~Takahashi, S.~Miyasaka, M.~Kibune, and H.~Takagi,
\emph{Spin-charge separation in single-chain compound Sr$_2$CuO$_3$ studied by	angle-resolved photoemission},
\href{http://dx.doi.org/10.1016/S0038-1098(98)00099-4}{Solid State Commun. {\bf 106}, 543 (1998)}.

\bibitem{Lorenzana97}
J.~Lorenzana and R.~Eder,
\emph{Dynamics of the one-dimensional Heisenberg model and optical absorption	of spinons in cuprate antiferromagnetic chains},
\href{http://dx.doi.org/10.1103/PhysRevB.55.R3358}{Phys. Rev. B {\bf 55}, R3358 (1997)}.

\bibitem{Suzuura96}
H.~Suzuura, H.~Yasuhara, A.~Furusaki, N.~Nagaosa, and Y.~Tokura,
\emph{Singularities in Optical Spectra of Quantum Spin Chains},
\href{http://dx.doi.org/10.1103/PhysRevLett.76.2579}{Phys. Rev. Lett. {\bf 76}, 2579 (1996)}.

\bibitem{Takigawa97}
M.~Takigawa, O.~A.~Starykh, A.~W.~Sandvik, and R.~R.~P.~Singh,
\emph{Nuclear relaxation in the spin-$1/2$	antiferromagnetic chain compound Sr$_2$CuO$_3$: Comparison between theories and experiments},
\href{http://dx.doi.org/10.1103/PhysRevB.56.13681}{Phys. Rev. B {\bf 56}, 13681 (1997)}.

\bibitem{Takigawa96}
M.~Takigawa, N.~Motoyama, H.~Eisaki, and S.~Uchida,
\emph{Dynamics in the $S=1/2$ One-Dimensional Antiferromagnet Sr$_2$CuO$_3$ via $^{63}$Cu NMR},
\href{http://dx.doi.org/10.1103/PhysRevLett.76.4612}{Phys. Rev. Lett. {\bf 76}, 4612 (1996)}.









\end{thebibliography}

\end{document}